\documentclass[floatfix,aps,prl,superscriptaddress, amsfonts]{revtex4-1}

\usepackage{graphicx}

\usepackage{dcolumn}

\usepackage{bm}
\usepackage{epstopdf}

\begin{document}
\title{One-Dimensional Alignment of Nanoparticles Via Magnetic Sorting}
\author{R. Bouskila}
\affiliation{Department of Physics \& Institute of Optical Sciences, University of Toronto, 60 St. George Street, Toronto, ON M5S 1A7}
\author{R. McAloney}
\affiliation{Department of Chemistry \& Institute of Optical Sciences, University of Toronto, 80 St. George Street, Toronto, ON M5S 3H6}
\author{S. Mack}
\affiliation{Center for Spintronics \& Quantum Computation, University of California, Santa Barbara, CA 93106}
\author{D.D. Awschalom}
\affiliation{Center for Spintronics \& Quantum Computation, University of California, Santa Barbara, CA 93106}
\author{M.C. Goh}
\affiliation{Department of Chemistry \& Institute of Optical Sciences, University of Toronto, 80 St. George Street, Toronto, ON M5S 3H6}
\author{K.S. Burch}
\email{kburch@physics.utoronto.ca}
\affiliation{Department of Physics \& Institute of Optical Sciences, University of Toronto, 60 St. George Street, Toronto, ON M5S 1A7}

\begin{abstract}
Near room temperature, thin films of MnAs spontaneously align into two phases, one ferromagnetic and the other paramagnetic. These two phases take the intriguing form of nanoscale "wires." In this experiment, we investigate the spontaneous formation of ordered linear arrays of magnetite nanoparticles on MnAs thin films using scanning probe microscopy.\end{abstract}

\maketitle
Many research groups have recently turned their attention to the problem of forming linear patterns of nanoscale particles \cite{Li,Lopes,Warner,Lopesdiblock,Aurongzeb,Engtrakul,Xu}. Such arrays are of interest for studies of the effects of low dimensionality on the electronic, magnetic and optical properties of materials, as well as for applications such as biological and chemical sensing \cite{xia, Horak}. The dominant paradigm is to fabricate a template, which nanoparticles will subsequently self-assemble along. This method, although successful, tends to be an expensive, complex process.\cite{xia} In this study, we have exploited the naturally-occuring magnetic order of MnAs thin films as a template for forming linear arrays of magnetite (Fe$_{3}$O$_{4}$) nanoparticles.

Bulk manganese arsenide was first investigated systematically by Guillaud \cite{Guillaud}. It is ferromagnetic at low temperatures (in the $\alpha$ phase), whereas at higher temperatures (above approximately 318 K) it undergoes a first-order structural phase transition, to the $\beta$ phase. Upon entering the $\beta$ phase MnAs experiences a reduction in the a-axis lattice constant and a spontaneous loss of magnetization occurs\cite{Menyuk}. For the last decade, there has been substantial interest in MnAs, as thin films can be grown on GaAs and Si substrates via molecular beam epitaxy (MBE).\cite{Gallas,Tanaka,Ney,Jenichen} These films, originally developed in the context of the microelectronics industry for applications in spintronics and magnetic storage on a chip, behave quite differently from the bulk material. As the film cools from its formation temperature, it expands in crossing the $\beta$\---$\alpha$ transition, experiencing compressive strain in the axis of expansion. This strain causes the film to form alternating linear domains of ferromagnetic $\alpha$ phase and paramagnetic $\beta$ phase. These domains are almost invisible to a topographical study such as atomic force microscopy (AFM), but show up in high relief in a magnetic force microscopy (MFM) scan\cite{Tanaka}. As detailed below, we have successfully exploited these linear domains to trap magnetite nanoparticles into one-dimensional arrays. This is confirmed via careful analysis of combined MFM and AFM studies.

\begin{figure}
 \includegraphics[width=6in]{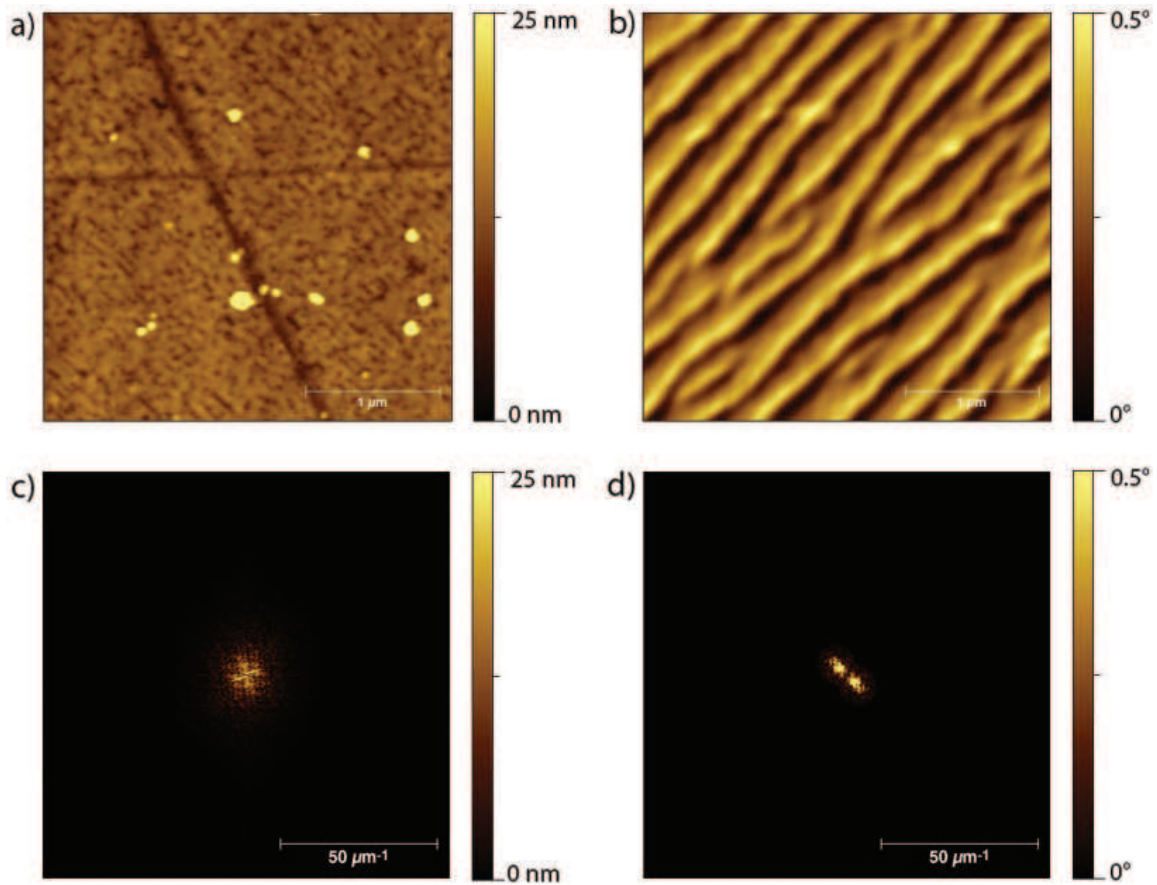}
 \caption[Bare MnAs]%
 {A bare MnAs film grown on a GaAs substrate. a) Topography of the film at room temperature exhibiting an RMS surface roughness of 2.31 nm, b) Magnetic field gradient measured in tandem with the AFM data; Bottom row: c) MFM of the data in a), revealing no evidence for order, d) Fourier transform of the data in b), clearly revealing two peaks due to the stripes.}
 \label{fgr:fig1}
\end{figure}

A 60 nm film of MnAs was epitaxially grown on a semi-insulating GaAs(001) substrate, as detailed in reference \citenum{Tanaka}. In the left panel of \ref{fgr:fig1} we present a room temperature AFM scan of our film before deposition of the nanoparticles. It is quite flat (RMS surface roughness of 2.31 nm), aside from some small particulates on the surface of the film, and two scratches crossing it. Furthermore, the fast fourier transform (FFT) of the AFM data reveals no evidence of ordering peaks. Confirming our assertion that the AFM is not sensitive to the phase separation. Turning to the MFM image shown in the right panel of \ref{fgr:fig1}, we show the change in the phase of the cantelever motion. This change in phase results from the force exerted on the cantilever by the stray magnetic fields of the film. In this image the one-dimensional alignment of the $\alpha$ and $\beta$ phases is clear, and is consistent with previous studies\cite{Ney,Jenichen}. The FFT of the MFM data shown in \ref{fgr:fig1}d reveals two clear peaks due to the phase separation, as discussed later. For the most part the MFM data is unrelated to the topography, except for some distortion around dust particles on the surface. This indicates that we can separately evaluate the existence of additional particles on the surface from the underlying magnetic ordering of the MnAs film. 

Colloidal nanoparticles of magnetite and titania (TiO2), chosen as a non-magnetic control) were provided by Vive Nano, Inc. These nanoparticles are 10 nm in diameter, and are surrounded with a small sphere of remanent colloid \cite{Anderson}. The particles were prepared in a solution of 0.01 M sodium hydroxide to prevent the formation of aggregates. These solutions were sonicated for 30 seconds to further break up any aggregates. Next a single drop of the nanoparticle solution was deposited on the MnAs film. As MnAs is a hydrophobic surface, the solution formed a bead upon deposition. A 30-second settling period was allowed so that the nanoparticles would find a local energy minimum via Brownian motion. Finally, the bulk of the solution was removed by wicking with laboratory wiping paper, leaving a thin film behind, which was subsequently allowed to dry.

After deposition, the samples were imaged using a Digital Instruments scanning probe device. We used magnetized Veeco (model MESP) MFM probes, which allowed us to perform magnetic force measurements interleaved with atomic force imaging \cite{sarid}. The TappingMode$^{\texttrademark}$ atomic force image revealed the surface topography and the rest positions of the settled nanoparticles, whereas the magnetic force image, taken at a lift height of 30 nm, revealed the orientation and location of the underlying MnAs magnetic domains. 

\begin{figure}
 \includegraphics[width=6in]{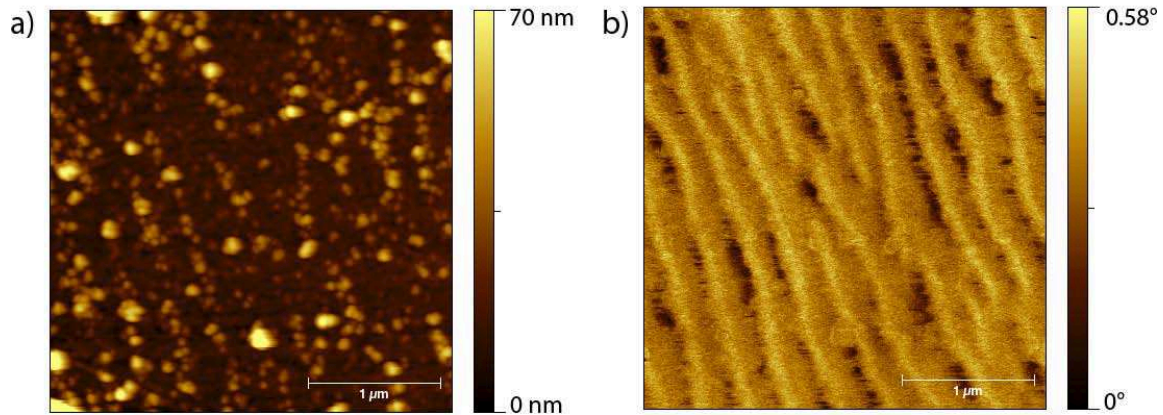}
 \caption[4$\mu$m$\times4\mu$m scans of magnetite nanoparticles on MnAs film.]%
 {4$\mu$m$\times4\mu$m scans of magnetite nanoparticles on MnAs film. a) AFM data showing the one-dimensional alignment of the nanoparticles, b) MFM 
data demonstrating that the alignment of the nanoparticles follows the underlying magnetic order.}
 \label{fgr:fig2}
\end{figure}

The nanoparticles appear in the AFM micrograph in \ref{fgr:fig2}a as approximately 10 nm increases in the height of the surface, with a circular diameter of approximately 50 nm. There is an apparent linear alignment of the particles. It bears note that although the diameter of the actual particles is very small, upon deposition their colloidal coating flattens out to a much wider diameter. This is evident in the circular diameter of the particles seen in the AFM micrographs. The existence of this colloid also makes it difficult to determine the exact position of the nanoparticle. Nonetheless, it is worth comparing the AFM data shown in \ref{fgr:fig2}a with the underlying magnetic domain structure as shown in the MFM image \ref{fgr:fig2}b. A cursory examination of these two figures clearly shows the direction of alignment of the nanoparticles closely follows the orientation of the nanowires in the film. 

\begin{figure}
 \includegraphics[width=6in]{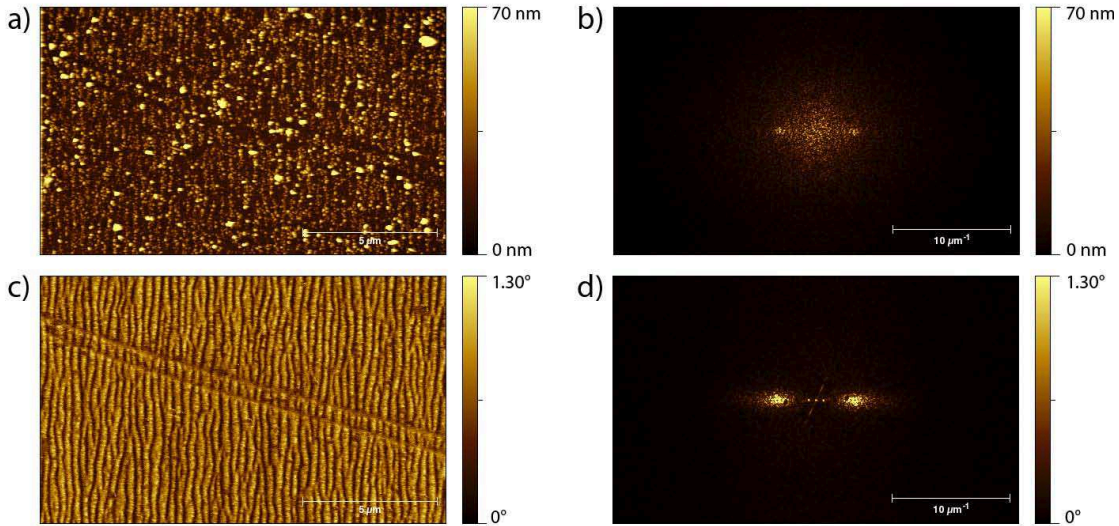}
 \caption[Magnetite nanoparticles on MnAs, sample 2]%
 {Magnetite nanoparticles on MnAs, sample 2. a) AFM, showing one dimensional nanoparticle alignment, with b) Fourier transform of the data presented in a. The alignment of the nanoparticles results in two peaks in the FFT; Bottom row: c) MFM, showing underlying MnAs striping, with d) the Fourier transform of c. The peaks in the FFT of the MFM match those of the AFM confirming the dots line up along with the magnetic order of the film.}
 \label{fgr:fig4}
\end{figure}

Interestingly, we have found this alignment exists over very large distances. This assertion is validated in \ref{fgr:fig4}, where we present a second sample, with a much wider scan size (10$\mu$m$\times15\mu$m). To confirm the alignment is the same for the Fe$_3$O$_4$ nanoparticles and MnAs nanowires we exploit the wealth of information it provided by the two-dimensional discrete fast Fourier transforms (FFTs). Indeed, from the FFT we can determine the periodicity, direction and correlation length of the magnetic domains and lines of nanoparticles. We wish to quantifiably determine whether or not the periodicity and direction of the lines of nanoparticles is in fact the same as that of the magnetic domains.

The FFTs of the AFM and MFM data presented in \ref{fgr:fig4}a \& \ref{fgr:fig4}b can be seen in \ref{fgr:fig4}c and \ref{fgr:fig4}d respectively. For linearly ordered patterns, the Fourier transform includes two peaks whose centers are equally spaced from q=0, and which lie along the axis of modulation (i.e., transverse to the lines). These peaks are clearly visible for the magnetic domains in MnAs (cf. \ref{fgr:fig4}d), but for a more complex structure, additional features can arise making the peaks more difficult to detect. Nevertheless, one can still see corresponding modulation peaks in both Fourier transform images. Indeed, both \ref{fgr:fig4}c and \ref{fgr:fig4}d include two bright Fourier peaks at q=$\pm3.1\mu$m$^{-1}$, indicating that the direction and period of the linearly ordered particles closely matches the underlying magnetic order of the MnAs substrate.

\begin{figure}
 \includegraphics[width=6in]{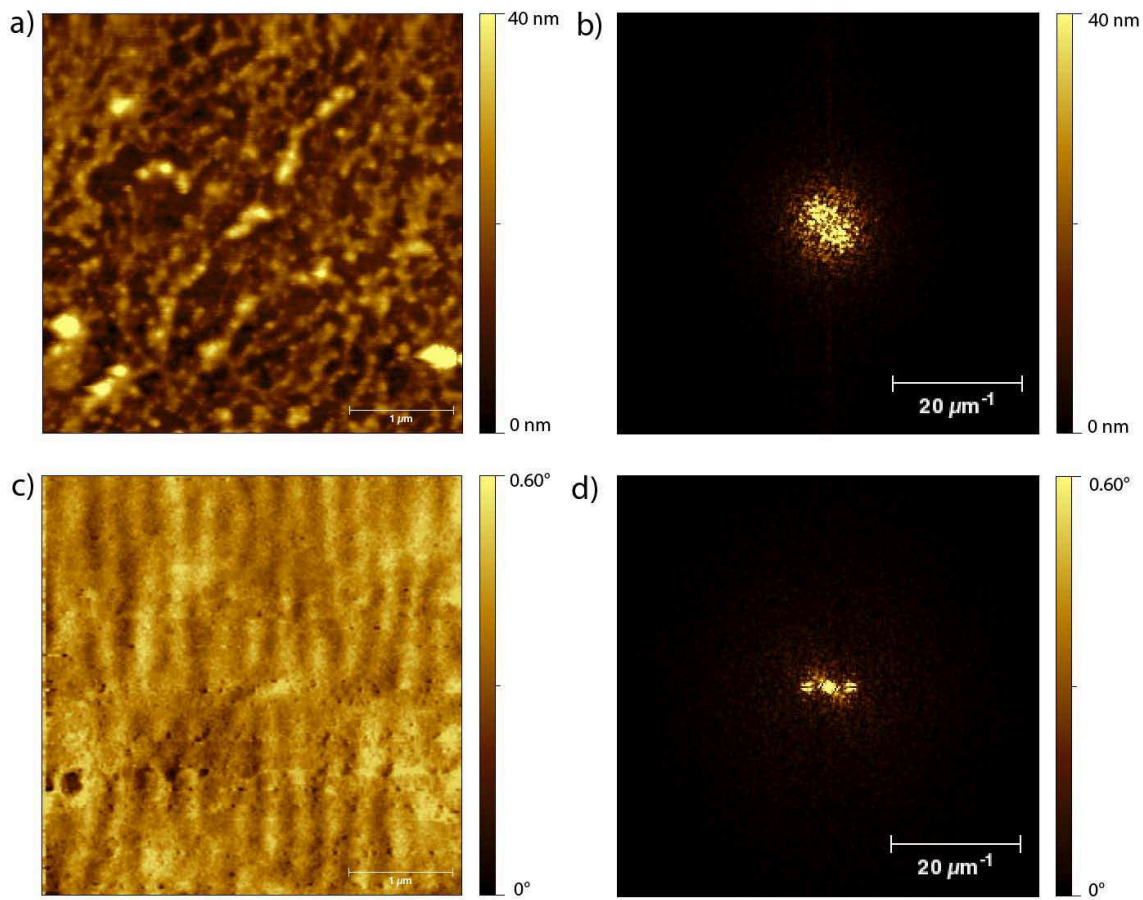}
 \caption[Titania nanoparticles on MnAs]%
 {Titania nanoparticles on MnAs. Top row: a) AFM, providing no evidence for TiO2 alignment, with b) Fourier transform of the data presented in a. Bottom row: c) MFM, showing underlying MnAs striping, with d) the Fourier transform of c. The MFM images taken together show this film also had magnetic order that did not effect the deposition of the TiO2 nanoparticles.}
 \label{fgr:fig5}
\end{figure}

To confirm that the alignment is due to the underlying magnetic order, the process described above was repeated using nonmagnetic titania nanoparticles. The results of this experiment are presented in \ref{fgr:fig5}. The MFM image (\ref{fgr:fig5}c) displays a vertical ordering, which is not apparent in the AFM image(\ref{fgr:fig5}a). The unordered state of the titania control sample is a strong indication that the ordering seen with magnetite particles is due to the high magnetic permeability of the magnetite particles, rather than due to the surface topography.

In \ref{tbl:table1} we tabulate some of the data from the Fourier transforms in \ref{fgr:fig4}.
\begin{table}
	\caption{Data extracted from the 2-D Fourier transform.}
	\label{tbl:table1}
	\begin{tabular}{lllllll}
		\hline %
		Sample & qpeak ($\mu$m$^{-1}$) & period ($\mu$m) & $\Gamma_1$ ($\mu$m$^{-1}$) & $\Gamma_2$ ($\mu$m$^{-1}$) & Lcorr$_1$ ($\mu$m) & Lcorr$_2$ ($\mu$m) \\
		\hline
		Magnetite, MFM & 3.10 & 0.322 & 1.32 & 0.397 & 0.758 & 2.51 \\
		Magnetite, AFM & 3.11 & 0.321 & 1.45 & 0.361 & 0.690 & 2.77 \\
		\hline
	\end{tabular}
\end{table}

We have evaluated the results presented in \ref{fgr:fig4}b and \ref{fgr:fig4}d by fitting each peak with two Guassians of width $\Gamma_1$(width along the ordering vector) and  $\Gamma_2$ (width normal to the ordering vector). The results of these fits are presented in Table 1.The FWHM of the Gaussian in the direction of the line between the two peaks ($\Gamma_1$), gives a measure of the variance in the spacing of the stripes, whereas $\Gamma_2$ gives a measure of the variance in the length of the stripes. These are simply the inverses of the correlation lengths in the direction of modulation (ie., over what distance the lines keep the same modulation period) and in the perpendicular direction (i.e., how long the stripes tend to be). The results presented in Table 1 indicate the period and correlation lengths for the MFM and AFM data are essentially the same. This result and the lack of ordering for the TiO$_{2}$ particles (see \ref{fgr:fig5}), strongly suggest the linear alignment of the Fe$_3$O$_4$ nanoparticles results from the underlying MnAs magnetic stripes.

\begin{figure}
 \includegraphics[width=6in]{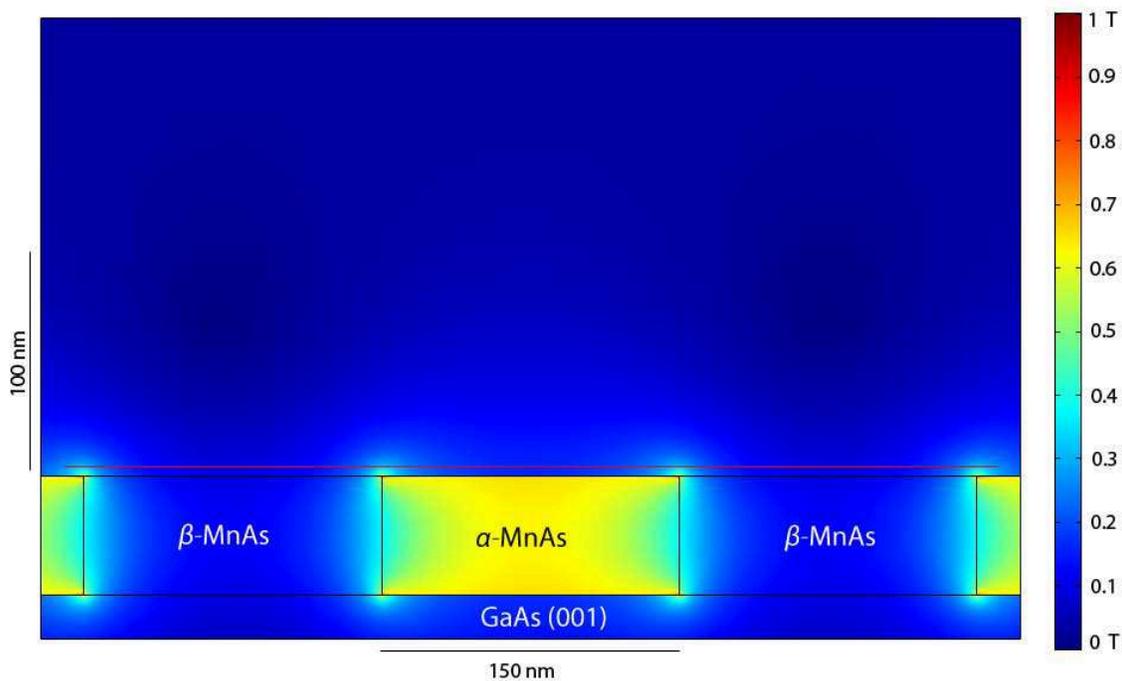}
 \caption[Magnetic field strength]%
 {Finite-element method simulation of the magnetic field strength in MnAs. The red line 5 nm above the MnAs surface indicates the average center position of the magnetite nanoparticles.}
 \label{fgr:simulation}
\end{figure}

Before concluding we discuss the mechanism by which the magnetite nanoparticles are aligned into one dimensional wires. Preliminary SQUID measurements of the Fe$_{3}$O$_{4}$ particles indicate they are superparamagnetic at room temperature with a specific magnetization ($\sigma$) of $4-10$ emu/g. The potential energy (U) for these particles in the field produced by the MnAs film ($\overrightarrow{B}$) is $U=-\overrightarrow{m}\cdot\overrightarrow{B}$. Since our experiments are performed with the particles are in solution at room temperature, we expect the direction of the magnetic moment ($\overrightarrow{m}$) to be easily rotated. Therefore so long as the magnetic field direction changes slowly compared with the average velocity of the particles, we expect the direction of $\overrightarrow{m}$ to closely follow that of $\overrightarrow{B}$. As a result, the particles should be trapped in regions with a high magnetic field. To determine the strength of the magnetic field ($|\overrightarrow{B}|$) we conducted a finite-element method simulation (via the software COMSOL Multiphysics) of the magnetic fields present in and around a thin film of MnAs. The results of these simulations for the strength of the field are shown in \ref{fgr:simulation}.

From the results presented above, it is apparent that the locations where the magnetic field is highest lie at the domain walls between the $\alpha$ and $\beta$ domains. To determine whether the potential generated by this field is strong enough to trap the partices, we treat them as point like particles with total $|\overrightarrow{m}|=(\frac{4\pi}{3})r^{3}\rho\sigma$; where r is the radius (5 nm) and $\rho$ the density of the nanoparticles (taken to be the same as bulk Fe$_3$O$_4$: 5.15~g/ cm$^{3}$).  Next we take the value $|\overrightarrow{B}|=0.3$T at a height of 5 nm above the surface (the radius of the magnetic nanoparticles). Putting this together, we find the trapping potential for the magnetic nanoparticles to be: $U=-|\overrightarrow{m}||\overrightarrow{B}|=780$~K $\rightarrow 1950$~K, where this range originates from the variety of magnetization values of the Fe$_3$O$_4$ particles. The rather large values of the confinement potential confirm that it is the magnetic fields of the MnAs nanowires that align the magnetite nanoparticles.  

A number of applications and further research possibilities present themselves for this method. The ability to trap magnetic particles and order them into linear arrays presents opportunities for self-assembling more complex molecules into such patterns, by functionalizing the nanoparticles with organic ligands. Since the MnAs domain widths can be controlled via temperature, one could impose a thermal gradient on the sample prior to deposition and thereby sort the nanoparticles by size. Lastly we note that the periodicity of the domains can be tuned with film thickness offering a wide degree of freedom for the nanoparticle alignment.

Another application emerging from this work is the detection of underlying magnetic order via deposition of magnetite nanoparticles on a substrate. Indeed we have demonstrated that these particles will align themselves with the underlying magnetic patterns, which could then be revealed by a simple AFM scan. Perhaps the most promising direction of research, however, would be to investigate whether one could deposit magnetite nanoparticles upon a substrate with a pre-designed magnetic field arrangement and thereby rapidly self-assemble arrays of nanoparticles in arbitrary patterns, without the need for a prefabricated mechanical template.

As a first step to these goals, we have demonstrated the formation of ordered linear patterns of magnetite nanoparticles upon an MnAs thin film substrate. It is apparent that the pattern formation relies upon the magnetic nature of the nanoparticles and the underlying magnetic order of the substrate. A method for the self-assembly of patterns of magnetic nanoparticles using nanoscale magnetic fields was proposed.

We are grateful for numerous discussions with A. Dattlebaum and G. Montano. Work at the University of Toronto was supported by NSERC, CFI, ORF and OCE. Work at UCSB was supported by DARPA and ONR.

\end{document}